# The Distributions in Nature and Entropy Principle

Oded Kafri

The derivation of the maximum entropy distribution of particles in boxes yields two kinds of distributions: a "bell-like" distribution and a long-tail distribution. The first one is obtained when the ratio between particles and boxes is low, and the second one - when the ratio is high. The obtained long tail distribution yields correctly the empirical Zipf law, Pareto's 20:80 rule and Benford's law. Therefore, it is concluded that the long tail and the "bell-like" distributions are outcomes of the tendency of statistical systems to maximize entropy.

**Introduction:**

There are two common distributions in life: The first one is the "bell-like" distribution, which is found in the distribution of IQ, human heights, human age at death etc. This "almost universal" distribution was introduced for the first time by Moivre in the 18$^{th}$ century and explored by Laplace and Gauss around 1800.

As opposed to the bell curve distribution, many quantities are distributed unevenly [1]. For example, the probability to live in a big city is higher than the probability to live in a small village. Similarly, the probability to be poor is higher than the probability to be rich. Although intuitively it is logical for cities' population and wealth to have a bell curve distribution, it is not so. Their distributions are uneven and are characterized by a long tail to the right, in which few have a lot and many have quite a little. These distributions were observed by Pareto, Zipf, Newcomb and Benford about a century later and received their name accordingly: Zipf law [2,3], Pareto's rule [4,5], and Benford's law [6,7].

The first to discover it was Pareto. In 1896 he observed that the ownership of lands in Italy is distributed among the population in the ratio of around 20:80, namely, about 20% of the population own about 80% of the land. From his observations of other countries as well, he concluded that this ratio is general. Mussolini embraced the Italian Marquis Pareto because he believed that the Pareto's rule proves nature's preference of the fittest. Zipf - a Harvard professor of linguistic - found out that the ratio between the first most frequent word and the second one, in any text in many languages, is two. Similarly, the ratio between the second most frequent word and the fourth one is also two, etc. He claimed that the shortest and most "efficient" words appear more frequently [2].

Zipf believed in the evolutionary philosophy, i.e. the most "useful" and "efficient" words are the winners, in the spirit of "the survival of the fittest". On the other hand, many people and political movements believe that Pareto's rule is unfair and the wealth should be shared more equally, namely, as in the bell curve distribution. The discovery of Newcomb about the uneven frequency of digits in logarithmic table in 1881 [6], (the higher the value of a digit, the lower its frequency) raises some doubts as for the real reason for the uneven distributions. Later, in 1938, Benford confirmed Newcomb's uneven distribution of digits in a wide range of numerical data [7]. He attempted, unsuccessfully, to present a formal proof to Newcomb's equation, see Eq.(12). Since than, this distribution was found also in prime numbers [8], physical constants, Fibonacci numbers and many more [9].

In this paper it is argued that the "bell-like" distribution and the long tail distribution are the boundaries of the same probability distribution. This probability function is obtained by a fair and unbiased random distribution of particles in boxes.

We consider a set of $N$ boxes scoring $P$ particles; it is assumed that all the boxes have an equal probability to score a particle, namely, the probability of a box to score a particle is $q = \frac{1}{N}$. Therefore, the probability to score $n$ particles is $q_n = \left(\frac{1}{N}\right)^n$. It is clear that $q_n < q$. This is the basic reason why the rich are fewer than the poor. In the case of $P \ll N$, where a multiple score in negligible, the "bell-like" distribution is obtained; and in the case of $P \gg N$, a long tail distribution is obtained.

### I. How $P$ particles are distributed in $N$ boxes?

The answer to it is not new: the particles are distributed in a way that maximizes the entropy [10].

According to Boltzmann, entropy is proportional to the maximum possible number of the different configurations (microstates) of a set. Namely,

$$S = \ln \Omega \qquad (1)$$

(we take here the Boltzmann constant $k_B \equiv 1$). A microstate is one possible distinguishable configuration of a set of boxes and particles. Boltzmann entropy is obtained from the Gibbs-Shannon entropy by assuming that all the microstates have an equal probability. The Gibbs-Shannon entropy is given by:

$$S = -\sum_{j=1}^{\Omega} p_j \ln p_j, \qquad (2)$$

where $p_j$ is the probability of the microstate $j$ and $\Omega$ is the number of microstates to be maximized. If all the microstates have an equal probability, namely, $p_j = \frac{1}{\Omega}$, Boltzmann entropy $\ln \Omega$ is obtained.

Therefore, the distribution of particles that maximizes Boltzmann entropy means an equal probability to any configuration as well as an equal probability to any particle to be in any box.

The number of microstates (different configurations) of $P$ particles in $N$ states is given by the Plank expression [10] namely,

$$\Omega(P,N) = \frac{(N+P-1)!}{P!(N-1)!}. \qquad (3)$$

To visualize the problem we start with a numerical example; namely, calculating the distribution of 3 particles in 3 boxes that maximizes entropy. According to Eq. (3) the number of microstates $\Omega(3,3)$ is 10, as follows:

3|0|0, 0|3|0, 0|0|3, 2|1|0, 2|0|1, 1|2|0, 0|2|1, 1|0|2, 0|1|2, and 1|1|1.

We see that although each box has an equal chance to score 1, 2, or 3 particles, the boxes with 1 particle appear 9 times, those with 2 particles appear 6 times, and those with 3 particles appear 3 times. The relative frequency of the boxes with one particle in a set of three boxes is therefore $f(1) = 0.5$; with two particles $f(2) = 0.33\bar{3}$ and with three particles $f(3) = 0.16\bar{6}$.

To calculate the relative frequencies $f(n)$, we designate $n = \dfrac{P}{N}$, where $n$ is the number of particles in a box, and apply the Stirling's formula $\ln N! \cong N \ln N - N$. We obtain [10] from Eqs.(1) and (3) that,

$$S \cong N\{(1+n)\ln(1+n) - n\ln n\} \cong \sum_{n=1}^{N}\{(1+n)\ln(1+n) - n\ln n\} \qquad (4)$$

Now we write the Lagrange equation,

$$F(n) \cong \sum_{n=1}^{N}\{(1+n)\ln(1+n) - n\ln n\} - \beta\{P - \sum_{n=1}^{N} n\phi(n)\} \qquad (5)$$

The first term on the RHS is the entropy and the second term is the constraint of the number of particles. Namely, $P = \sum_{n=1}^{N} n\phi(n)$ is the number of particles, $\phi(n)$ is the number of boxes that scored $n$ particles and $\beta$ is a Lagrange multiplier. $\phi(n)$ can be interpreted as the probability of a box to have $n$ particles. The normalized $\phi(n)$, $f(n)$ is the relative frequency of the boxes that scored $n$ particles. From $\frac{\partial F(n)}{\partial n} = 0$ one obtains,

$$\phi(n) = \beta^{-1} \ln(1 + \frac{1}{n}) \qquad (6)$$

Eq.(6) is the analogue of Planck equation, [11,12,13] namely,

$$n = \frac{1}{e^{\beta\phi(n)} - 1}. \qquad (7)$$

Hereafter, we examine three cases:

In the first case we assume that $n \gg 1$. Here one can expect to find a large number of particles (limited by $P$) in any of the boxes. For example, if we conduct a popularity poll between the $N$ words among $P$ authors, and there are many more authors than words, then the maximum entropy distribution of the votes between the words is shown to be the Zipf law.

In the second case we consider the intermediate zone where $n$ is in the range of the number of the boxes. This case fits well to the distribution of ranks, namely, Pareto's rule and Benford's law.

In the third case we consider $n \ll 1$, where the number of particles is negligible as compared to the number of boxes. This case fits well to the probability of guessing correctly the IQ of a person in a single guess based only on the knowledge of the average. This case yields the "bell-like" distribution.

**IIa Zipf law:** Consider the case where $P \gg N$ where $n \gg 1$. In this case $\beta\phi \ll 1$, therefore from Eq.(7) $\phi(n)$ can be approximated to,

$$n\phi(n) \cong \frac{1}{\beta} \qquad (8)$$

Eq.(8) is the Zipf law. Namely, the ratio in the frequencies between $n=1$ (the most frequent word) and $n=2$ (the second most frequently word) is 2 which is identical to the ratio between $n=2$ and $n=4$ etc. This ratio is not a function of $\beta$ as, $\dfrac{\phi(1)}{\phi(2)} = \dfrac{\phi(2)}{\phi(4)} = \ldots = \dfrac{\phi(n)}{\phi(2n)} \cong 2$.

**IIb Pareto's rule:** to calculate the relative frequency of Eq.(6), namely, $f(n)$ we have to divide $\phi(n)$ by the sum over all the $M$ occupied boxes $M \leq N$, namely,

$$\sum_{i=1}^{M} \phi(n) = \beta^{-1}(\ln\frac{2}{1} + \ln\frac{3}{2} + \ldots + \ln\frac{M+1}{M}) = \beta^{-1}\ln(M+1). \qquad (9)$$

Therefore,

$$f(n) = \frac{\ln(1+\frac{1}{n})}{\ln(M+1)} \qquad (10)$$

Like in the Zipf law, for integer $n$'s, the relative frequency is not a function of $\beta$. We define a rank $r \equiv n\frac{N}{P}$ where $r = 1,2,3,......,R$. By defining the ranks we combined the boxes into clusters of boxes such that each cluster will contain $r = 1,2,3,......,R$ groups of $\frac{P}{N}$ particles. $r = 10$ means 10 times more particles than $r = 1$. We can repeat the calculation of the frequency again but instead of using $n$, we will use $r$, and obtain;

$$f(r) = \frac{\ln(1+\frac{1}{r})}{\ln(R+1)} \qquad (11)$$

In Fig.1 The relative frequencies for a set of $R = 10^6$ clusters and $r = 1,2,3,....,R$ according to Eq.(11) is plotted. A long tail distribution is demonstrated.

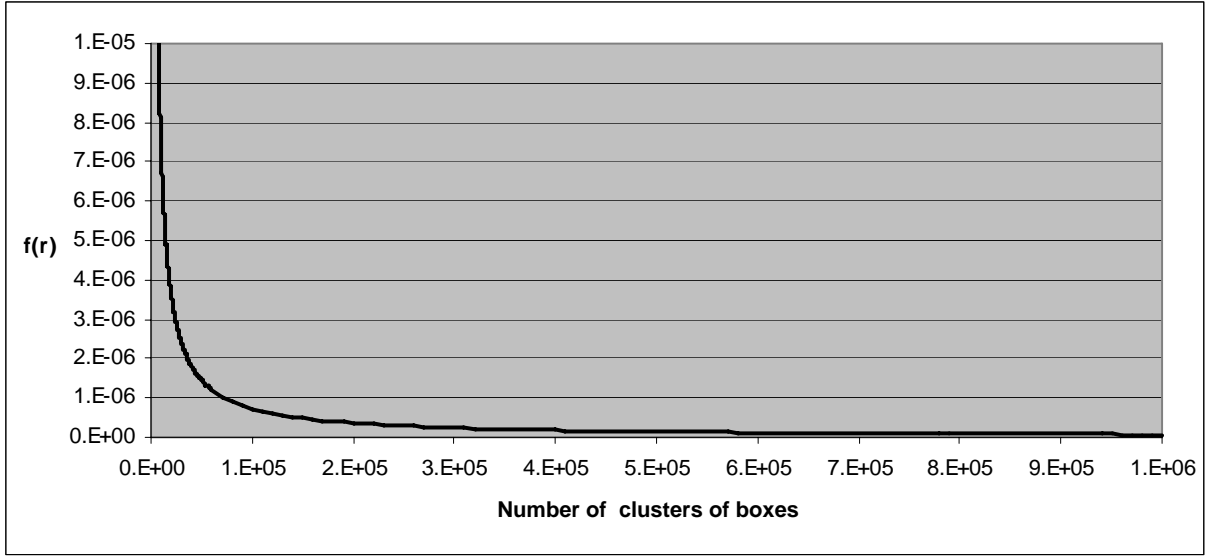

Fig.1: *A million clusters and their probabilities. The rank increases as its probability decreases.*

Eq.(11) "behaves" as a power law, this is so because a plot of the logarithm of the cluster *r* versus the logarithm of its probability yields a straight line as demonstrated for a million ranks.

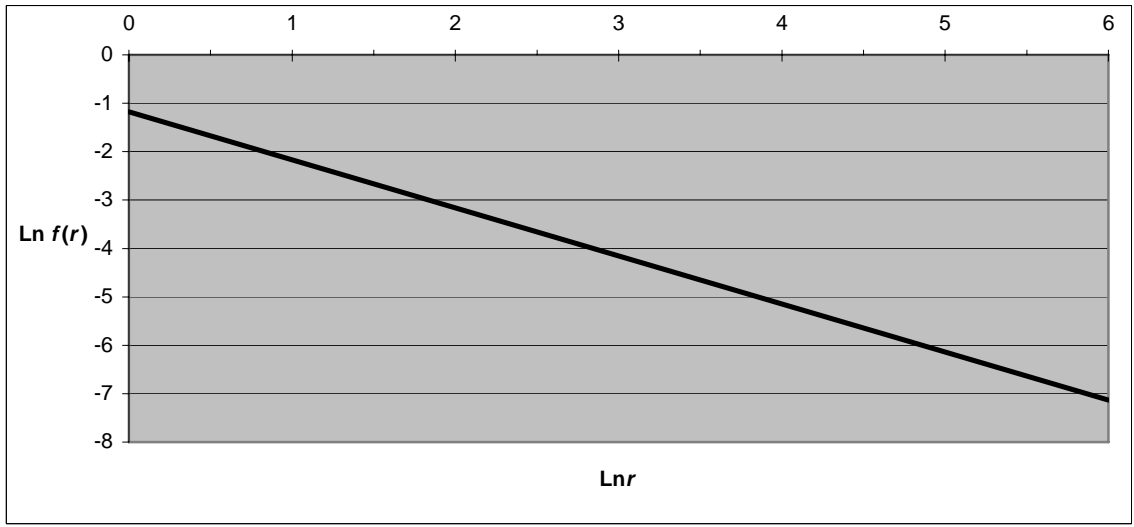

Fig.2: *Log-Log plot of frequency versus the rank for R=million is a straight line.*

The Pareto's 20:80 rule of thumb was proved to be correct not only in wealth distribution but in many other phenomena as well. For example, it is believed that 20 percent of customers yields 80 percent of the revenue; 20 percent of the drivers cause 80 percent of the accidents; etc [5]. In order to find the ratio obtained from Eq.(11) we divide the boxes into 10 ranks. Each rank contains 1, 2, 3,….,9, 10 equal groups of particles. We construct the table below from $f(r) = \dfrac{\ln(1+\frac{1}{r})}{\ln 11}$ :

| r | 10 | 9 | 8 | 7 | 6 | 5 | 4 | 3 | 2 | 1 |
|---|----|---|---|---|---|---|---|---|---|---|
| f(r)% | 4 | 4.4 | 4.9 | 5.6 | 6.6 | 7.6 | 9.3 | 12 | 16.9 | 28.9 |

*Table 1: The relative frequencies of* 10 ranks

The total number of groups is $\sum_{r=1}^{10} r = 55$. However, the richest five ranks contain $\sum_{r=6}^{10} r = 40$ groups. Their total frequencies are $\sum_{r=6}^{10} f(r) = 25.5\%$, which means that about 73% of the packages are in the hands of about 25% of the boxes. This is a typical behavior of the Pareto's rule but with a small deviation from the empirical rule of thumb of 20:80, namely, a 25:75 rule.

**IIc Benford's Law**: Another application of Eq.(11) is Benford's law. Newcomb suggested Benford's law in 1881 from observations of the physical tear and wear of books containing logarithmic tables [6]. Benford further explored the phenomenon in 1938, and empirically checked it for a wide range of numerical data. The main application of Benford's distribution is based on its existence in numerous random numerical files like financial data, street addresses, etc. Since one intuitively expects to obtain an even distribution of digits, as would be in the case of an unbiased lottery, some income tax authorities are looking at balance sheets for digit distributions in order to detect fraud detection. If the balance sheets don't fit to Benford's law, a further inspection is done [14].

In the derivation of Benford's law we assume that a digit is a box with $n$ particles. This assumption is logical as a digit, unlike a word, has an absolute meaning as compared to other digits, exactly as the meaning of the number of particles in a box. There is a constraint though: the number of particles in a digit cannot exceed 9. The digit zero does not appear in Benford's law distribution of the first order. In Eq.(11) $r$ may have any number. In digits, per definition, $r \leq 9$, therefore, it is legitimate to calculate the equilibrium distribution of the occupied boxes and to add as many empty boxes without affecting the distribution. In this case $R$ is 9 and Eq.(11) yields the relative frequency,

$$f(r) = \frac{\ln(1+\frac{1}{r})}{\ln 10} = \log_{10}(1+\frac{1}{r}) \qquad (12)$$

This is the Benford's law.

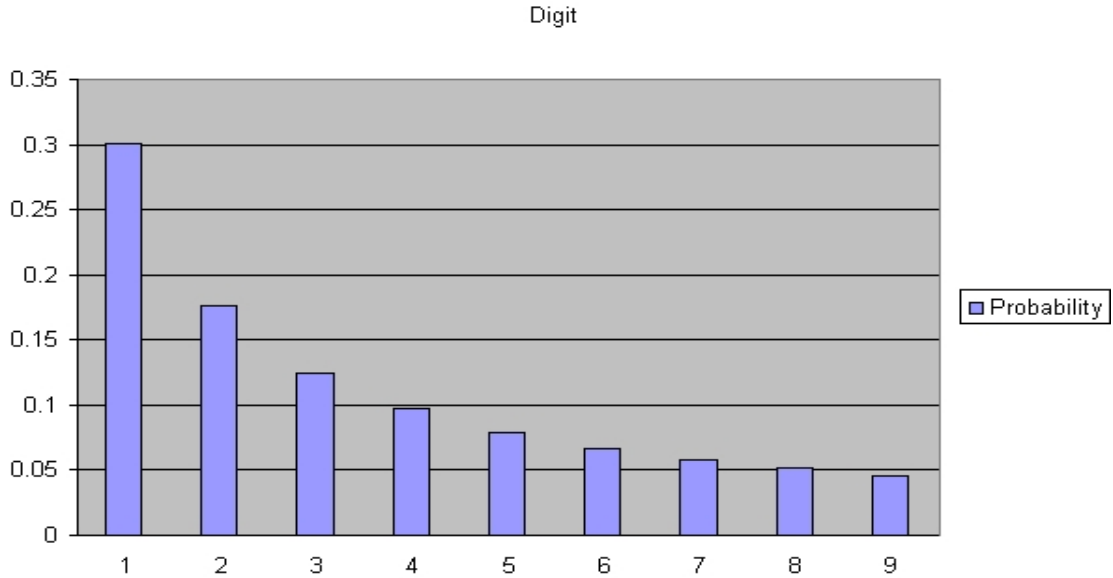

Fig 3. *Benford's law predicts a decreasing frequency of first digits, from 1 through 9.*

**III "bell-like" distribution:** Zipf law, Pareto's rule and Benford's law occurs where the number of particles is larger than the number of boxes. Hereafter, the case where $P \ll N$ is considered.

In the case where $n \ll 1$, we neglect the boxes that scored several particles, because, practically there are no such boxes. We want to find the probability distribution of $N$ boxes to score one particle. In this limit, $e^{\beta\phi} \gg 1$ and Eq.(7) can be approximated to,

$$n_i = e^{-\beta\phi_i} \qquad (13)$$

Here $n_i$ is the fraction of a particle in a box and the frequency $\phi_i = \phi(n_i)$ is the probability to find this fraction. The total number of particles $P$ is given by the same expression that we used in the Lagrange equation (5) namely,

$$P = N\phi_i n_i = N\phi_i e^{-\beta\phi_i}, \qquad (14)$$

in the limit $\beta \to 0$ one obtains that all the frequencies $\phi_i$ of the boxes are equal, namely $\phi = \frac{P}{N}$. This is an even distribution. The even distribution is the intuitive distribution that one expects to find in a distribution of particles in boxes. This distribution causes us to believe that uneven distributions are counterintuitive.

In the case where $\beta$ is finite

$$P = \sum_{i=1}^{N} \phi_i e^{-\beta\phi_i} = \sum_{i=1}^{N} P(\phi_i, \beta). \qquad (15)$$

$\frac{P(\phi_i, \beta)}{P}$ is the relative probability to find a particle in a box. From Eq.(15) it is seen that $P(\phi_i, \beta)$ has two components, the first is the frequency $\phi_i$ of the fraction $n_i$ of the particle and the second is the fraction of particles. As opposed to the case where $P \gg N$, the frequency $\phi(n)$ itself is not the probability to find *n* particles but the probability to find a fraction of a particle. To find the probability of a single particle we have to multiply the frequency by the fraction of the particle namely $\phi_i n_i$. When the frequency increases the associate fraction of particles decreases exponentially with the frequency. The larger the $\beta$, the steeper is the decay. Since $P(\phi, \beta)$ is a linearly increasing function of $\phi_i$ multiplied by an exponentially decay function of $\phi_i$, the distribution of particles in a box has a definite maximum.

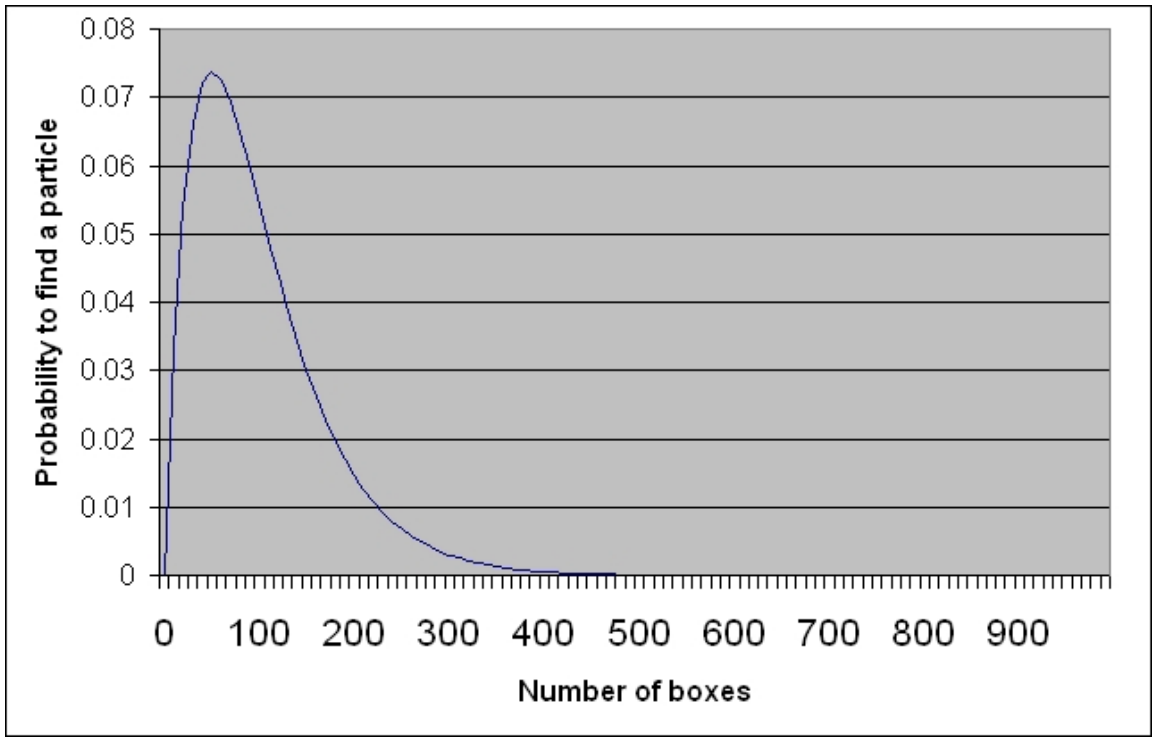

Fig 4: *The number of boxes, and their probability to find a single particle for N=1000 and $\beta = \frac{1}{50}$.*

The maximum probability is obtained from $\frac{\partial P}{\partial \phi} = e^{-\beta\phi} - \beta\phi e^{-\beta\phi} = 0$ and is given by $\phi_{max} = \frac{1}{\beta}$. In Fig.(4) we see that the obtained curve is typical of velocity of molecules, human age at death etc.

**IV Discussion**: The long tail distribution attracts a considerable attention because it is so ubiquitous [15]. Sometimes it is called a power law distribution and scale-free distribution. This is because a Log-Log presentation of the distribution yields a straight line as seen in Fig.2. When a power law fits are done, different slopes obtained for different statistics. For example, in Zipf law the ratio between the frequency of the 1st and the frequency of the 2nd is 2; in Pareto's rule and in Benford's law this ratio is about 1.7. Namely, in different regimes of $P/N$ different "slopes" are obtained as is seen in Fig. 5. Another notable point is that the normalized frequencies $f(n)$ for $P \gg N$ are not a function of $\beta$. This is with contradistinction to the case $P \ll N$ in which the distribution is a function of $\beta$.

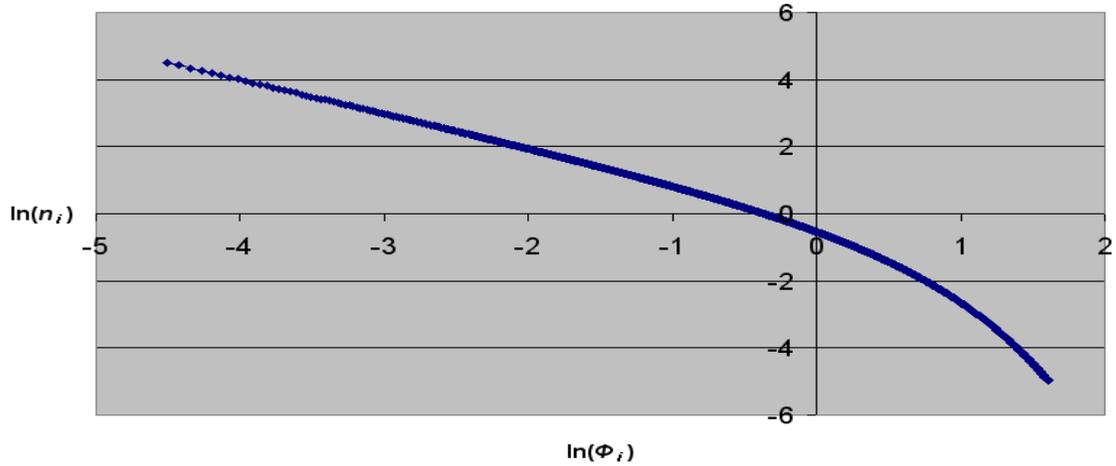

Fig.5. *a plot of* $\ln\phi$ *versus* $\ln n$ *: for high values of n a "power law" decay is obtained, however for low values of n an exponential decay is obtained.*

The Lagrange multiplier $\beta$ has a meaning. In thermodynamic the temperature is related to it via $T \propto \frac{1}{\beta}$. We see that in the case of Zipf law the frequency multiplied by the number of particles is proportional to the temperature. In the case of $n \ll 1$ the temperature is proportional to the frequency in which the probability to find a particle is the highest. This is the main difference between the long tail distribution and the "bell-like" distribution. In the long tail the temperature means the average wealth of a box. In the bell curve the temperature means the average maximum probability.

**Summary:** The distribution of $P$ non-interacting particles in $N$ boxes is calculated for a fair system. Since there is no preference to any configuration of particles and boxes, the entropy principle can be applied. It is shown that when the number of the particles is negligible as compared to the number of boxes, the "bell-like" distribution (which prefers the average) is obtained. However, when the number of particles is higher than the number of boxes, a long tail distribution is obtained. The obtained long tail distribution yields correctly Zipf law, Pareto's rule and Benford's law.

The Pareto's rule usually is conceived as an evolutionary law. Namely, the 20% of the drivers that cause 80% of the accidents are the bad drivers. Maybe the personality of these drivers is the reason for their excessive involvement in car accidents. Similarly, there might be good reasons for the fact that few people get rich and the majority remains poor. These kinds of questions cannot be answered by this kind of analysis. However, one should bear in mind that particles without personality, interactions or statistical bias are also distributed in the same way.

**Acknowledgment:** I acknowledge Alex Ely Kossovsky, Dan Weiss and H. Kafri for most valuable comments.